\documentclass[sigconf]{acmart}

\newcommand{\sparagraph}[1]{\vspace{1mm}\noindent {\bf #1}}

\usepackage{multirow}
\usepackage{array}
\usepackage{balance}
\usepackage{pifont}
\newcommand{\cmark}{\ding{51}}
\newcommand{\xmark}{\ding{55}}

\AtBeginDocument{%
  }

\copyrightyear{2025}
\acmYear{2025}
\setcopyright{cc}
\setcctype{by}
\acmConference[aiDM '25]{International Workshop on Exploiting Artificial Intelligence Techniques for Data Management}{June 22--27, 2025}{Berlin, Germany}
\acmBooktitle{International Workshop on Exploiting Artificial Intelligence Techniques for Data Management (aiDM '25), June 22--27, 2025, Berlin, Germany}
\acmDOI{10.1145/3735403.3735998}
\acmISBN{979-8-4007-1920-2/2025/06}

\title{Redbench: A Benchmark Reflecting Real Workloads}

\author{Skander Krid}
\orcid{0009-0007-2055-2012}
\affiliation{%
  \institution{University of Technology Nuremberg}
  \city{Nuremberg}
  \country{Germany}
}
\email{skander.krid@utn.de}

\author{Mihail Stoian}
\orcid{0000-0002-8843-3374}
\affiliation{%
  \institution{University of Technology Nuremberg}
  \city{Nuremberg}
  \country{Germany}
}
\email{mihail.stoian@utn.de}

\author{Andreas Kipf}
\orcid{0000-0003-3463-0564}
\affiliation{%
  \institution{University of Technology Nuremberg}
  \city{Nuremberg}
  \country{Germany}
}
\email{andreas.kipf@utn.de}

\renewcommand{\shortauthors}{Krid et al.}

\begin{document}

\begin{abstract}
    Instance-optimized components have made their way into production systems. To some extent, this adoption is due to the characteristics of customer workloads, which can be individually leveraged during the model training phase. However, there is a gap between research and industry that impedes the development of realistic learned components: the lack of suitable workloads. Existing ones, such as TPC-H and TPC-DS, and even more recent ones, such as DSB and CAB, fail to exhibit real workload patterns, particularly distribution shifts.
    In this paper, we introduce Redbench, a collection of 30 workloads that reflect query patterns observed in the real world. The workloads were obtained by sampling queries from support benchmarks and aligning them with workload characteristics observed in Redset.
\end{abstract}



\maketitle

\section{Introduction}

In one of his BTW 2025 talks, Goetz Graefe made a distinctive statement: ``We, as a community, have very few ideas.'' We feel compelled to add: ``but when that happens, that idea changes the way we build and optimize systems.''
A notable example is the shift from the traditional approach of optimizing databases for worst-case scenarios to instance-optimized systems, i.e., components that adapt to the characteristics of the observed workloads~\cite{kraska}.
This new paradigm is well suited to production environments where user workloads exhibit diverse patterns making one-size-fits-all worst-case optimizations suboptimal.

\sparagraph{Motivation.} However, this naturally incurs a gap between research and industry: While production systems have the possibility to stress test their workload-driven optimizations, e.g., by performing a workload replay, the research community is often constrained to using benchmarks that do not accurately reflect query patterns seen in practice. In particular, workload distribution shifts determine the difference between a robust and an impractical optimization.
Without efforts in this direction, the impact of upcoming learned or workload-driven optimizations will be limited.

\sparagraph{Research $\Join$ Industry.} Fortunately, industry has been open to this call and released in the last years statistics about production workloads. For instance, Snowset~\cite{snowset} and Redset~\cite{redset} capture customer query metadata from Snowflake and Redshift, respectively. To this end, van Renen and Leis~\cite{cab} analyzed Snowset and proposed the Cloud-Analytics Benchmark (CAB), which integrates the findings found in Snowset.
CAB does use TPC-H as its support benchmark, yet it incorporates essential cloud-specific dimensions, such as elasticity and multi-tenancy. Since the release of Snowset, there has been an increased interest in leveraging the \emph{repetitiveness} of queries and table scan expressions for clustering~\cite{mddl}, secondary indexing~\cite{pred_cache}, query super-optimization~\cite{super_marcus}, and resource allocation~\cite{intelligent_scaling1, intelligent_scalling2}.
Notably, repetitiveness is a key feature that workload-driven optimizations rely on and is not well captured by any of the standard benchmarks, as shown by van Renen et al.~\cite{redset}, nor by CAB, which randomly samples queries from the support benchmark. DSB~\cite{dsb} aims to address this issue, particularly in TPC-DS. Yet, the underlying workload statistics are generated synthetically.

\sparagraph{Contribution.} In light of recent advancements in the learned systems community, we outline the characteristics a benchmark should fulfill to properly stress-test such proposals.
Then, we introduce Redbench, a benchmark that maps a support benchmark to Redset by preserving these very characteristics that proved to be crucial in the workload-driven optimizations of the last few years. Redbench's workloads for well-studied support benchmarks are available at: \texttt{https://github.com/utndatasystems/redbench}.

\sparagraph{Structure.} The paper is structured as follows: We first provide an overview of recent learned and workload-driven optimizations, extracting the characteristics that they require from a benchmark to be properly tested (Sec.~\ref{sec:components}).
Then, we introduce Redbench's sampling and mapping mechanisms (Sec.~\ref{sec:redbench}).
Finally, we outline future work and plans for upcoming releases of Redbench (Sec.~\ref{sec:conclusion}).

\section{Instance-Optimized Components}\label{sec:components}

\begin{table*}[h]
    \caption{What learned and workload-aware optimizations require from a benchmark for proper validation.}
    \label{tab:all}
    \centering
    \begin{tabular}{|c|p{3.75cm}|c|c|c|c|}
        \hline
        \textbf{Problem} & \parbox{4.0cm}{\centering \textbf{Instance-Optimized Components}} & \textbf{What is needed} & \textbf{TPC} & \textbf{DSB} & \textbf{Redbench} \\
        \hline
            \multirow{3}{*}{Caching} &
                Semantic Caching~\cite{semantic_data_caching1}: Crystal~\cite{crystal}, Cache Investment~\cite{cache_investment},\newline
                Plan Caching~\cite{plan_caching1, plan_caching2} &
            \multirow{3}{*}{query rep.} & \multirow{3}{*}{\xmark} & \multirow{3}{*}{fixed} & \multirow{3}{*}{\cmark} \\
        \hline
        \multirow{2}{*}{Clustering (incl.~Secondary Indexing)} & MDDL~\cite{mddl},&
            \multicolumn{1}{c|}{table scan rep.} & \xmark & fixed & v0.4 \\
        \cline{3-6} 
        &  Predicate Caching~\cite{pred_cache} & workload drift & \xmark & synthetic & real \\
        \hline
            \multirow{4}{*}{Query Optimization} &
                Learned Cardinalities~\cite{learned_cards1, learned_cards2},
                Neo~\cite{neo},
                Bao~\cite{bao},
                Balsa~\cite{balsa} &
           \multirow{2}{*}{query rep.} &  \multirow{2}{*}{\xmark} &  \multirow{2}{*}{fixed} & \multirow{2}{*}{\cmark} \\
            \cline{3-6}
            & 
                Lero~\cite{lero}, FlowLoss~\cite{ceb},\newline
                Super-Optimization~\cite{super_marcus}
            & \multirow{2}{*}{workload drift} & \multirow{2}{*}{\xmark} & \multirow{2}{*}{synthetic} & \multirow{2}{*}{real} \\
        \hline
        \multirow{2}{*}{Materialized Views~\cite{mv1, mv2, mv3}} & DynaMat~\cite{dynamat}, Hawc~\cite{hawc}, CloudViews~\cite{cloud_views} & \multirow{2}{*}{query rep.} & \multirow{2}{*}{\xmark} & \multirow{2}{*}{fixed} & \multirow{2}{*}{\cmark} \\
        \hline
        Resource Allocation (incl.~Query Stats Prediction) & Intelligent Scaling~\cite{intelligent_scaling1, intelligent_scalling2} & \parbox{2.25cm}{\centering timestamp-based\\query rep.} & \xmark & \xmark & v0.5 \\
        \hline
    \end{tabular}
\end{table*}

Real-world workloads exhibit patterns that system components optimizing for the worst-case scenarios do not take advantage of. Learned and workload-aware components exploit this very fact by modeling data and/or workload distribution during training and data structure building.

\newpage

In recent years, our community has spent a significant amount of time optimizing these proposals. Even so, the benchmarks used for evaluation have lagged behind, since they do not capture the query patterns seen in production systems. To understand this gap, consider Tab.~\ref{tab:all},
where we examine (1) the standard problems in databases, (2) existing learned/workload-aware components, (3) what these components require from a benchmark, and (4) whether existing benchmarks actually meet these requirements.
Before we outline the categorization made in column (3) of the same table, we briefly review related work.

\sparagraph{Existing Benchmarks.} TPC-benchmarks such as TPC-H~\cite{tpch} and TPC-DS~\cite{tpcds} are widely used to test both research and production systems, yet they operate on synthetic data. JOB~\cite{job}, which aims to stress test the query optimizer, runs on the real-world IMDb dataset, yet its queries are handwritten. An extension of JOB is CEB~\cite{cab}, which provides more JOB-like queries on the same IMDb dataset. Neither of the aforementioned benchmarks accurately reflects the characteristics of production workloads. DSB~\cite{dsb} is a recent proposal that builds on TPC-DS, by providing more complex data distributions and dynamic workloads. However, the query repetition ratio is fixed, and the workload drifts are generated randomly using a Gaussian distribution.

\sparagraph{Clustering.} In order to reduce I/O, data systems engineers spend a considerable amount of effort designing pruning strategies for data blocks~\cite{snowflake-paper}.
This is not surprising, as queries spend around 50\% of their time in scan and filter operators~\cite{cab}. Notably, this is also the case for read-only queries, where scan and filter operators account for 44.5\% of the total execution time.
Traditionally, this pruning effect is enabled by clustering the data. However, this decision can be rather arbitrary if the workload accesses multiple columns.

A promising direction is to allow the system to analyze the characteristics of the workload and adapt accordingly. For example, MDDL~\cite{mddl} exploits the repetitiveness of table predicates to sort on multiple columns at once. Notably, it can converge to the optimal number of data blocks, provided that all user predicates are captured within its auxiliary column. The challenge arises in the presence of a workload drift, e.g., the set of table predicates changes completely, rendering the auxiliary column unusable. We plan to focus on table scan repetitiveness as part of Redbench v0.4, particularly on matching the findings in Schmidt et al.~\cite[Fig.~5]{pred_cache}.

\sparagraph{Caching.} Workload drifts also limit the applicability of workload-driven secondary indexes such as Predicate Caching~\cite{pred_cache}, which maintains a lightweight key-value cache of all table and semi-join-induced predicates on base tables. The values in the cache are approximations of the row ranges that qualify for the filter. Workload drifts affect the relevance of these cache entries. Having a workload-aware benchmark opens the possibility for testing predicate-relaxation techniques~\cite{pred_cache_relaxed}, e.g., rows qualifying for \texttt{o\_orderdate <= '2025-05-01'} also qualify for later months.

\sparagraph{Query Optimization.} Given the repetitiveness of queries, a recent proposal by Marcus~\cite{super_marcus} envisions long optimization times instead of the traditional, fast optimization strategies that only consider statistics.
The intuition is that it is worth spending more time optimizing recurring queries, the overhead being amortized over the subsequent runs.
Evaluating such proposals requires realistic query repetition patterns.

\sparagraph{View Maintenance.} Materialized Views (MVs) are a complementary approach to optimize query performance by pre-computing subresults. An important decision around MVs is which views to choose. The goal is to achieve the greatest performance benefit relative to the associated maintenance and storage overhead.
Leveraging workload information---particularly the repetitiveness of joins, predicates, and aggregations---can simplify this decision.

\sparagraph{Resource Allocation.} In a serverless environment, a system must be able to handle workload spikes~\cite{cab}.
As a result, accurately predicting the timestamps---possibly even the queries---associated with peak periods is highly valuable.
The same is true for resource allocation, where predicting query execution time and memory consumption can further improve the decision on which resources to use during these bursts. Redbench v0.5 will include timestamps, allowing it to be integrated with CAB (Redbench can become one of CAB's query pools).

Next, we present a benchmark designed to bridge this gap by capturing all key characteristics of real-world analytical workloads discussed so far.
\newpage

\section{Redbench}\label{sec:redbench}

\begin{figure}[]
  \centering
  \includegraphics[width=1.0\linewidth, keepaspectratio]{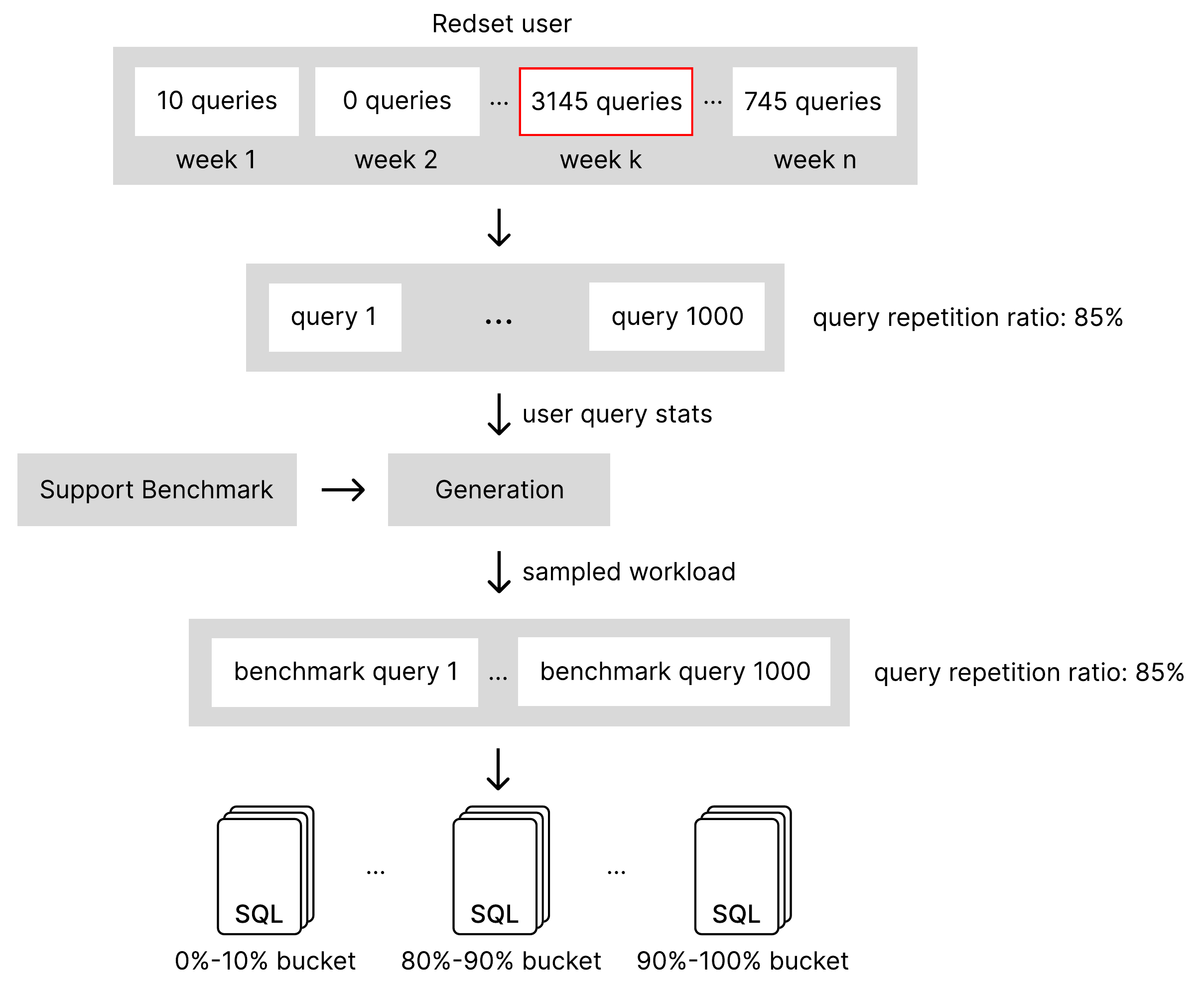}
  \caption{Sampling and generation using a support benchmark. Redbench's workloads are partitioned into 10 buckets, corresponding to the query repetition ratio in Redset~\cite{redset}.}
  \label{fig:workload_sampling}
\end{figure}

We introduce Redbench, a repetition-aware benchmark based on Redset, the customer query metadata published by Redshift~\cite{redset} spanning 3 months and 200 clusters. The idea behind Redbench is to sample users with different workload profiles from Redset, and recreate their workloads by sampling similar queries from a support benchmark. The sampling process of a single user workload is visualized in Fig.~\ref{fig:workload_sampling}. In the following, we detail the individual steps.

\subsection{Definitions}

For a Redset query, we define its \texttt{scanset} as the set of persistent tables that it scans, and its $\texttt{hash}$ as the combination of \texttt{scanset}, number of joins, number of table scans, and the $\texttt{feature\_fingerprint}$ field, which is a proxy for query likeness in Redset.

A key metric that we use to categorize a Redset user workload is the query repetition rate: the ratio of queries whose $\texttt{hash}$ has already occurred in the timeline. A user workload in Redset and its corresponding workload sampled by Redbench will have equal query repetition rates. We describe in Sec.~\ref{sec:sample_workload} how our workload sampling guarantees this.

\subsection{Prefiltering}

We prefilter Redset queries as follows:
\begin{itemize}
    \item Currently, Redbench supports only \texttt{SELECT}-queries, which account for 48.9\% of Redshift's fleet as shown in van Renen et al.~\cite{redset}. Support for updates is planned for v1.0.
    \item We filter out queries that were answered from Redshift's result cache. 
    A system with a result cache can answer such queries by just scanning the cached output. Therefore, we believe that these queries are not worth optimizing.
    \item During the workload sampling step, we will match Redset scansets with benchmark templates based on the number of joins. Thus, we eliminate Redset queries that do not contain any joins. Also since we choose the mapping based on the normalized number of joins, we eliminate users whose minimum and maximum numbers of joins are equal.
    \item For a consistent mapping, we want a scanset to uniquely identify the number of joins. For most queries, assuming no self-joins or set operators are present, the number of joins is equal to the scanset size minus one. We eliminate all Redset queries that do not adhere to this.
\end{itemize}

\sparagraph{Busiest Week.} For each user, we identify the week with the highest number of queries and retain only the first $K$ queries from that week; the value of $K$ is user-specified and currently set to 1,000. A week starts on Monday 8 am, and ends on Friday 5 pm.

\subsection{Support Benchmark} \label{sec:support-benchmarks}

Redbench is designed to be benchmark-agnostic, i.e., it does not make any assumptions on the support benchmark. However, it is crucial to make sure that the chosen benchmark has enough query variability, namely query templates and query instances, so it can accurately replicate various Redset's user workloads. A suitable support benchmark must also guarantee that query instances of the same template must have the same number of joins. This is the case for most well-studied benchmarks, e.g., JOB~\cite{job}, CEB~\cite{ceb}, TPC-H~\cite{tpch}, and TPC-DS~\cite{tpcds}. As of the current version, we provide:

\sparagraph{\texttt{redbench[imdb]}.} Redbench instantiation with the combination of the IMDb-based Join Order Benchmark (JOB) and Cardinality Estimation Benchmark (CEB). Combining both benchmarks provides a wider variety of query templates and instances to sample from.

\sparagraph{\texttt{redbench[tpc-ds]}.} Redbench instantiation with TPC-DS as the support benchmark. Given TPC-DS's large number of templates, it can be easily used as a support benchmark for Redbench.

\subsection{User Sampling}

We divide users into 10 buckets based on their Redset's query repetition rate: 0\%-10\%, 10\%-20\%, \ldots, 90\%-100\% (inclusive to the left). To ensure more variety in Redbench's workloads, we choose three users from each repetition bucket, for a total of 30 users, as follows:

\begin{itemize}
    \item[(a)] We rank users in each bucket in ascending order, once based on the number of distinct number of joins values, and once based on the number of distinct scansets across their queries.
    \item[(b)] We sum up both ranks for each user to get the user's workload-variability. A lower value means less diverse queries in terms of number of joins and scansets.
    \item[(c)] We select the users with the lowest, median, and highest workload-variability values.
\end{itemize}

\subsection{Workload Sampling}\label{sec:sample_workload}

In this step, we recreate each of the 30 user workloads from the previous step using queries from the support benchmark. Given a workload, we iterate over the queries in ascending order of their arrival timestamps, and map them to a benchmark query:

\begin{itemize}
    \item[(a)] If we have already encountered the query hash, reuse the same benchmark query instance mapped to it.
    \item[(b)] If we have already encountered the scanset, lookup the corresponding benchmark query template that we mapped to it and choose one of its unused query instances. We implement a fallback strategy in case we have already used up all query instances from the corresponding template (see below).
    \item[(c)] If we have not encountered the scanset before, then we need to map it to one of the unused templates of the support benchmark. We only consider templates that produce a normalized number of join closest to the normalized number of joins of the user query, and call these \emph{closest templates}. Then, we choose the template with the most number of query instances. It is possible that all templates have already been mapped. In that case, we rely on our fallback strategy.
\end{itemize}

\sparagraph{Fallback Strategy.} Reaching the fallback step means that the used support benchmark does not have enough query instances or templates to mimic the production workload. This happens for around 18.16\% and 17.33\% of all queries across the 30 workloads in \texttt{redbench[imdb]}  and \texttt{redbench[tpc-ds]}, respectively. We proceed as follows: If at least one of the already mapped benchmark closest templates has unused query instances, then randomly pick one of those. Otherwise, randomly pick and reuse one of the query instances from the closest templates.

\subsection{Workload Characteristics Preservation}

Redbench aims to retain key workload characteristics such as table scan repetitiveness and queries' relative complexity and timeline.

\begin{figure}[htpb]
    \centering
    \includegraphics[width=1\linewidth]{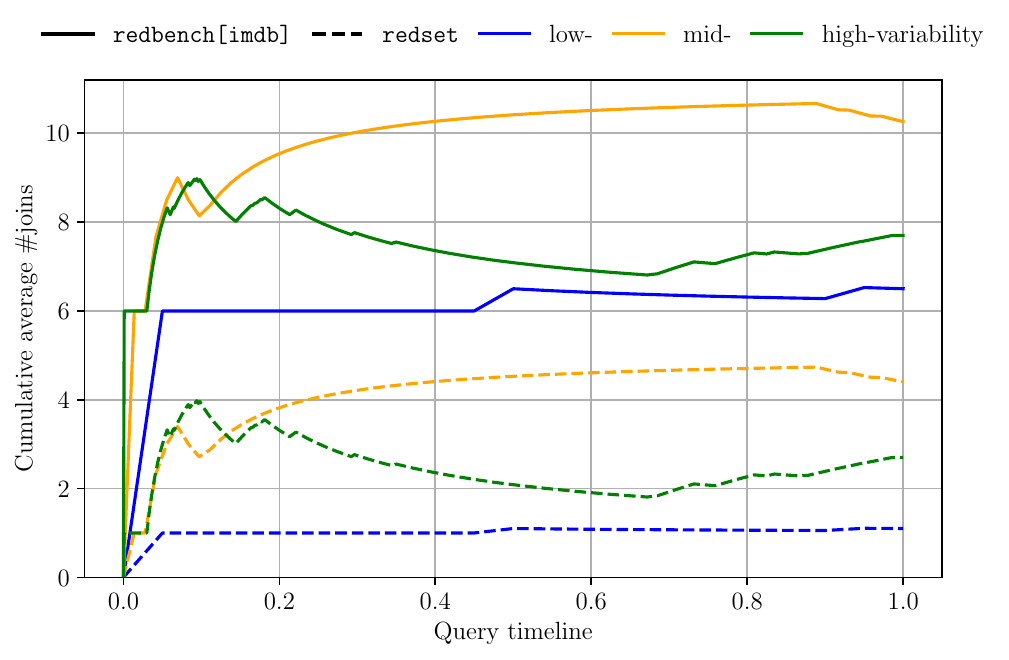}
    \caption{How Redbench's sampled workloads preserve Redset's relative join complexity over time for the query repetition bucket 80\%-90\%.}
    \label{fig:main-plot}
\end{figure}

\sparagraph{Number of Joins.} Currently, Redbench preserves the number of joins from Redset in a relative sense. Specifically, the number of joins in each query is normalized during mapping, i.e., the minimum and maximum number of joins in the Redset workload are linearly mapped to the minimum and maximum number of joins in the support benchmark. As shown in Fig.~\ref{fig:main-plot}, Redbench maintains the shape and structure of the join complexity curve over time despite the shifted absolute numbers.

\sparagraph{Scansets.} Ideally, Redbench would achieve a one-to-one mapping between user-level tables in Redset and tables in the support benchmark. In this scenario, each Redset query’s scanset would correspond uniquely to the scanset of a benchmark query, allowing precise replication of table access patterns and, by extension, join graph structures. However, two main limitations prevent this:
\begin{itemize}
    \item[(a)] Insufficient table coverage in support benchmarks: 31\% of all remaining Redset users after the prefiltering step query more tables than the number of tables available in either IMDb or TPC-DS, making a bijection impossible.
    \item[(b)] Limited query template diversity: Even when table and template counts match, if the user’s workload includes scansets like \texttt{[T1]} and \texttt{[T2]}, and the benchmark only includes \texttt{[T1]} and \texttt{[T1, T2]}, then no bijective mapping of user tables to benchmark tables can preserve the scanset structure.
    Therefore, a suitable support benchmark should ideally include a template for each possible scanset.
\end{itemize}

Instead, Redbench offers a best-effort preservation of scansets, bounded by the richness of the available support benchmark. In future versions of Redbench, we plan to consider synthetically generated queries as support benchmark. Assuming we can generate a query for each possible scanset, Redbench would achieve a perfect scanset-to-scanset equivalence with Redset.

\sparagraph{Full Query Repetition.} Another critical characteristic preserved by Redbench is full query repetition, i.e., the frequency and pattern with which identical queries reoccur in a workload. By construction, Redbench precisely replicates the query repetition rate observed in Redset by mapping identical query hashes to the same benchmark query instances. However, as mentioned earlier, we exclude Redset queries that were answered from the result cache.
Consequently, the remaining repetitions typically represent \texttt{SELECT} queries interspersed with updates, a common workload pattern found in business intelligence and dashboarding use cases \cite{redset}.

\sparagraph{Timeline and Workload Peaks.} Redbench currently preserves the relative temporal order of queries and plays them back-to-back. While sufficient to evaluate order but time-independent optimizations, discarding absolute timestamps is not suitable for real-time benchmarking scenarios. In a future version, we intend to integrate Redbench with CAB so that workload peaks can be accurately represented and used to evaluate workload-driven resource allocation strategies. CAB preserves wall‑clock timing and can drive realistic stress tests for elastic cloud systems.

\section{Conclusion}\label{sec:conclusion}

Redbench is a novel benchmark that reflects the query patterns present in production workloads. Being based on Redset~\cite{redset}, it inherits the natural properties that instance-optimized components require for a proper stress-test, such as query and table scan repetitiveness, and workload drifts. Redbench's agnosticity to the support benchmark makes it easy to instantiate it with an already existing one, provided it contains enough distinct query templates and query instances, to make it conform to Redset's statistics. Currently, Redbench has instantiations for TPC-DS~\cite{tpcds} and IMDb~\cite{job, ceb}.

\sparagraph{Future Work.} We plan to provide Redbench's instantiations of all well-studied benchmarks as query pools for CAB~\cite{cab}. Moreover, we plan to extend Redbench with updates so that instance-optimized components can be evaluated under realistic conditions.

\balance
\bibliographystyle{ACM-Reference-Format}
\bibliography{redbench}

\end{document}